\documentclass[12pt,aps,noshowpacs]{revtex4}%
\usepackage{amssymb}
\usepackage{amsmath}
\usepackage{amsfonts}%
\usepackage{graphicx}
\providecommand{\U}[1]{\protect\rule{.1in}{.1in}}
\begin{document}
\title{Acceleration and Classical Electromagnetic Radiation}
\author{E.N. Glass}
\affiliation{Department of Physics, University of Michigan, Ann Arbor, Michgan; Department
of Physics, University of Windsor, Windsor, Ontario}
\date{7 September 2007}

\begin{abstract}
Classical radiation from an accelerated charge is reviewed along with the
reciprocal topic of accelerated observers detecting radiation from a static
charge. \newline This review commemerates Bahram Mashhoon's 60$^{\text{th}}$
birthday.\newline\newline Keywords: Field theory, radiation, accelerating charges

\end{abstract}
\maketitle

\section{INTRODUCTION}

It is generally accepted by all physicists that when a charge is accelerated
in an inertial frame, observers in that frame detect the emission of
electromagnetic radiation. During the first half of the $20^{th}$ century,
measurements and theory combined to form a complete framework for this part of
classical electrodynamics. The radiated power of an accelerated charge is
expressed by the Larmor formula and goes as the square of the charge times the
square of the acceleration. A clear analysis leading to the Larmor formula is
given in Jackson's text \cite{Jac99} and in Rohrlich's text \cite{Roh65}, to
cite only two of many. Classical radiation from an accelerated charge is
reviewed in Part I.

Radiation can be observed when the relative acceleration between charge and
observer is non-zero. Part II reviews the issue of accelerated observers
detecting radiation from static charges. Mashhoon \cite{Mas90} has emphasized
the concept of locality and its role in measurements made by noninertial observers.

For a different view of radiation detected by accelerating observers, free
fall in a Reissner-Nordstr\"{o}m (RN) manifold is studied. Details of the RN
metric are presented in Appendix A, and a radial free fall frame is developed
in Appendix B.

\section{Part I}

The conformal symmetry of Maxwell's equations in Minkowski spacetime allows a
motionless Coulomb charge to be mapped to uniformly accelerated hyperbolic
motion. (For finite times this is a special case of general bounded
accelerations). This motion satisfies Rohrlich's local criterion \cite{Roh65}
for electromagnetic radiation which requires projecting the electromagnetic
stress-energy 4-momentum of a point charge into an inertial observer's rest
frame. The Lorentz scalar obtained by projection is then integrated over a
2-sphere whose radius can be much smaller than the radiation wavelength. The
integral provides the relativistic Larmor formula for radiated power (Gaussian
units, metric signature -2)
\begin{equation}
\mathcal{R}=-\frac{2}{3}\frac{q^{2}}{c^{3}}a_{\mu}a^{\mu} \label{larmor}%
\end{equation}
where $a^{\mu}$ is the 4-acceleration of charge $q$. An accelerating charge
satisfies this criterion and has non-zero $\mathcal{R}$.

In the following, a brief derivation of the field and energy-momentum of a
charge moving on a curved path in Minkowski spacetime is given.

\subsection{Radiative field}

Charge $q$ has 4-current%
\[
j^{\mu}(x)=q%
{\displaystyle\int\limits_{-\infty}^{+\infty}}
v^{\mu}(s)\ \delta^{(4)}[x-z(s)]ds
\]
where $s$ is the proper time along the worldline of $q$ with tangent $v^{\mu
}(s)=dz^{\mu}/ds$, $v^{\mu}v_{\mu}=c^{2}$ ($c=1$ hereafter).

Maxwell's equations (and the Lorenz gauge) provide%
\[
\square A^{\mu}(x)=4\pi j^{\mu}(x).
\]
Green's identity gives the retarded vector potential in terms of the 4-current%
\[
A_{\text{ret}}^{\mu}(x)=4\pi%
{\displaystyle\int\limits_{4-vol}}
D_{\text{ret}}(x-x^{\prime})j^{\mu}(x^{\prime})d^{4}x^{\prime}%
\]
(note that $\mu$ is a Cartesian index which passes through the integral).
$D_{\text{ret}}$ is the retarded Jordan-Pauli function \cite{Hei60}.%
\begin{equation}
D_{\text{ret}}(x-z)=\frac{1}{2\pi}\theta(\tau)\delta\lbrack(x^{\nu}-z^{\nu
})(x_{\nu}-z_{\nu})]\label{D-ret}%
\end{equation}
where $\tau=x^{0}-z^{0}$, with step function $\theta(\tau):=\{_{0\text{\ \ \ }%
\tau<0}^{1\text{ \ \ }\tau>0}$.
\begin{equation}
A_{\text{ret}}^{\mu}(x)=2q%
{\displaystyle\int\limits_{-\infty}^{+\infty}}
v^{\mu}(s)\delta\lbrack(x^{\nu}-z^{\nu})(x_{\nu}-z_{\nu})]ds.\label{A-1}%
\end{equation}
Since a forward null cone can emanate from each point on the worldline, we
label the worldline points with $u$, such that $s>u$. Let $f^{2}(s)$ be the
square of the spacelike distance between null cone field point $x^{\nu}$ and
worldline point $z^{\nu}$: $f^{2}(s)=-(x^{\mu}-z^{\mu})(x^{\nu}-z^{\nu}%
)\eta_{\mu\nu}$. It follows that%
\[
\lbrack2f\frac{df}{ds}]_{s=u}=2v_{\nu}(u)(x^{\nu}-z^{\nu}),
\]
which motivates the definition $R:=v_{\nu}(x^{\nu}-z^{\nu})$, the distance
along the null cone (the retarded distance between $x^{\nu}$ and $z^{\nu}$).
Note that, in the rest frame of the charge, $R=x^{0}-z^{0}$. From
Eq.(\ref{A-1}) we obtain%
\begin{equation}
A_{\text{ret}}^{\mu}(x)=2q%
{\displaystyle\int\limits_{-\infty}^{+\infty}}
v^{\mu}\text{ }\frac{\delta(s-u)}{2v_{\alpha}(u)(x^{\alpha}-z^{\alpha})}\text{
}ds=q\frac{v^{\mu}(u)}{R}\label{L-W}%
\end{equation}
the retarded Lienard-Wiechart potential.

The worldline is now parametrized by $u$, i.e. $z^{\alpha}(u)$. The distance
between points on the null cone is zero. $(x^{\alpha}-z^{\alpha})(x_{\alpha
}-z_{\alpha})=0$. The gradient $\partial/\partial x^{\mu}$ of the zero product
yields
\begin{align*}
(\delta_{\ \mu}^{\alpha}-\frac{dz^{\alpha}}{du}\partial_{\mu}u)(x_{\alpha
}-z_{\alpha})  &  =0\\
(\delta_{\ \mu}^{\alpha}-v^{\alpha}\partial_{\mu}u)(x_{\alpha}-z_{\alpha})  &
=0\\
(x_{\mu}-z_{\mu})-R\partial_{\mu}u  &  =0.
\end{align*}
The null vector $k_{\mu}$ along which the radiation propagates on the forward
light cone is defined as%
\begin{equation}
k_{\mu}:=\partial_{\mu}u=(x_{\mu}-z_{\mu})/R. \label{k-vec}%
\end{equation}
From the definition of $R$ it follows that
\[
\partial_{\mu}R=Rk_{\mu}(a_{\alpha}k^{\alpha})+v_{\mu}-k_{\mu}%
\]
where $a^{\alpha}:=dv^{\alpha}/du$ is the acceleration along the worldline. A
unit spacelike vector is defined as $\hat{n}_{\alpha}:=k_{\alpha}-v_{\alpha}$
such that%
\begin{align*}
\hat{n}_{\alpha}\hat{n}^{\alpha}  &  =-1,\text{ }\hat{n}_{\alpha}v^{\alpha
}=0,\text{ }\hat{n}_{\alpha}k^{\alpha}=-1,\text{ }k^{\alpha}v_{\alpha}=1,\\
v^{\alpha}v_{\alpha}  &  =1,\text{ }k_{\alpha}k^{\alpha}=0.
\end{align*}
It is convenient to define a curvature scalar $\kappa:=a_{\alpha}k^{\alpha}$.
The gradient of $R$ can be rewritten as%
\[
\partial_{\mu}R=\kappa Rk_{\mu}-\hat{n}_{\mu}.
\]
Differentiation of the Lienard-Wiechart potential (\ref{L-W}) provides
\[
A_{\text{ret}}^{\nu,\mu}(x)=\frac{q}{R}k^{\mu}(a^{\nu}-\kappa v^{\nu}%
)+\frac{q}{R^{2}}\hat{n}^{\mu}v^{\nu}.
\]
Define spacelike $y^{\nu}:=a^{\nu}-\kappa v^{\nu}$, orthogonal to $k^{\nu}$.
The electromagnetic field is%
\begin{align}
F_{\text{ret}}^{\mu\nu}  &  =2A_{\text{ret}}^{[\nu,\mu]}\nonumber\\
&  =2\frac{q}{R}k^{[\mu}y^{\nu]}+2\frac{q}{R^{2}}\hat{n}^{[\mu}v^{\nu]}
\label{F-ret}%
\end{align}
showing the $1/R$ radiation field and the $1/R^{2}$ velocity field
\cite{Tei70}.

\subsection{Energy-Momentum}

The symmetric energy-momentum tensor, for signature $[+,-,-,-]$, is%
\[
4\pi T^{\mu\nu}=-F^{\mu\alpha}F_{\ \alpha}^{\nu}+\frac{1}{4}\eta^{\mu\nu
}F^{\alpha\beta}F_{\alpha\beta}.
\]
The radiation part of $F_{\text{ret}}^{\mu\nu}$ has energy-momentum%
\begin{equation}
4\pi T^{\mu\nu}=\frac{q^{2}}{R^{2}}(y_{\alpha}y^{\alpha})k^{\mu}k^{\nu}.
\label{en-mom}%
\end{equation}

Consider a 4-volume bounded by two null cones $\mathcal{N}_{1}$ and
$\mathcal{N}_{2}$ and two timelike $R=const$ surfaces $\Sigma_{1}$ and
$\Sigma_{2}$ with normal $n^{\alpha}$. The limit $R\rightarrow\infty$ slides
$\Sigma_{1}$ and $\Sigma_{2}$ to null infinity while maintaining a finite
radial separation. Since $\partial_{\nu}T^{\mu\nu}=0$, integration of
$T^{\mu\nu}$ is cast onto the boundaries $\Sigma_{1}$ and $\Sigma_{2}$ at null
infinity. The volume between $\Sigma_{1}$ and $\Sigma_{2}$ is $R^{2}d\Omega
du$ (with $d\Omega=\sin\vartheta d\vartheta d\varphi$). The energy and
momentum radiated to null infinity in the interval between $\mathcal{N}_{1}$
and $\mathcal{N}_{2}$ is
\begin{align}
\frac{d\mathcal{E}^{\mu}}{du}  &  =\text{ }~_{R\rightarrow\infty}^{\text{lim}%
}\frac{1}{4\pi}%
{\displaystyle\oint}
T^{\mu\nu}n_{\nu}R^{2}d\Omega\nonumber\\
&  =-q^{2}(y_{\alpha}y^{\alpha})k^{\mu}=-q^{2}(a_{\alpha}a^{\alpha}+\kappa
^{2})k^{\mu} \label{en-loss}%
\end{align}
Since $k^{0}>0$, it follows that $d\mathcal{E}^{0}/du\geq0$ in all inertial
frames. Therefore $d\mathcal{E}^{\mu}/du$ is a timelike vector, which implies
there is no inertial observer for whom the total momentum decreases without a
corresponding energy loss.

\section{Part II}

Since an electric charge accelerating in an inertial frame emits
electromagnetic radiation, the reciprocal question is whether an accelerated
observer in an inertial frame, moving past a static charge (i.e. moving
through an electrostatic field) detects electromagnetic radiation? We review
the arguments for this question.

It is known that observers who accelerate through empty Minkowski space
observe a heat bath \cite{Unr76},\cite{Dav75}. This supports the notion that
such observers can see radiation in an electrostatic field, but this is a
quantum effect, outside the scope of this review.

Rohrlich \cite{Roh63}, in a study of the equivalence principle, asks and
answers the question "does radiation from a uniformly accelerated charge
contradict the equivalence principle?" The equivalence principle can be
defined as the existence of a local spacetime region where gravitational tidal
forces vanish, i.e. $R_{\mu\nu\alpha\beta}\delta x^{\mu}\delta x^{\nu}\delta
x^{\alpha}\delta x^{\beta}\sim0$. He then shows that in such a region, a
charge in hyperbolic motion satisfies a local criterion \cite{Roh65} for
radiation. Rohrlich further concludes that a uniformly accelerating detector
\textbf{will} absorb electromagnetic energy from a static charge. Fugmann and
Kretzschmar \cite{FK91} have generalized Rohrlich's results to arbitrary acceleration.

Mould \cite{Mou64} has provided invariant criteria to identify the absorption
characteristics of such detectors. Mould constructs a simple antenna and
computes the absorbed energy from a radiating charge on a hyperbolic
trajectory. He shows that a freely falling detector \textbf{will} absorb
energy from a static electric field.

Taken together, Rohrlich and Mould have fully answered the question of whether
accelerated observers detect radiation from static charges. In the following,
we affirm their answer from a different perspective.

\subsection{Free Fall}

A freely falling observer is attached to a family of local inertial reference
frames, isomorphic to the frames in which Maxwell's equations are expressed
and verified. \ Although Newtonian gravity is completely adequate for this
topic, we introduce the complexity of Einstein's gravitation in order to have
analytic expressions for observers freely falling in a static electric field
with non-uniform acceleration.

The RN metric $g^{\text{RN}}$ describes an electrostatic, spherically
symmetric, spacetime. Inside the RN trapped surface the source has parameters
$(m_{0},q)$, with $q$ the source of a static Coulomb field. The inertial frame
which exists at asymptotic spatial infinity can be used to span the RN
spacetime. We ask whether an accelerated observer in this spacetime can see
electromagnetic radiation?

Details of the RN spacetime are given in Appendix A.

\subsection{Maxwell field seen from free fall}

The Maxwell tensor for the RN Coulomb field, from Eq.(\ref{f_ten}), is%
\[
F_{\alpha\beta}^{\text{RN}}=-(\frac{q}{r^{2}})\ 2\hat{t}_{[\alpha}\hat
{r}_{\beta]}%
\]
with basis vectors which span the RN metric in curvature coordinates. Since
the RN\ spacetime is constructed with its source $(m_{0},q)$ a priori trapped,
an infalling observer cannot see the charge itself. The observer can only take
local samples of the surrounding static Coulomb field which can be projected
on a forward or backward light cone emanating from the observer's position.
Here we will use a forward cone. An expansion of $R_{q}(r)$ from
Eq.(\ref{R-q}) provides%
\[
R_{q}=\left(  \frac{m_{0}^{2}}{2q^{3}}\right)  r^{2}+O(r^{3}).
\]
Thus
\begin{equation}
r^{2}=2(\frac{q^{3}}{m_{0}^{2}})R_{q}+O(R_{q}^{2}). \label{r-R}%
\end{equation}
From Appendix B, we find the relationship between the static RN frame and the
radial free fall null frame%
\begin{align}
\hat{t}_{\alpha}  &  =[(1+A_{q}^{1/2})/2]L_{\alpha}+(1+A_{q}^{1/2}%
)^{-1}N_{\alpha},\\
\hat{r}_{\alpha}  &  =(1+A_{q}^{1/2})^{-1}N_{\alpha}-[(1+A_{q}^{1/2}%
)/2]L_{\alpha}.
\end{align}
It follows that
\[
2\hat{t}_{[\alpha}\hat{r}_{\beta]}=L_{[\alpha}N_{\beta]}.
\]
The free fall Maxwell tensor is
\begin{equation}
F_{\alpha\beta}^{\text{RN-ff}}=\left[  -\frac{1}{2}\frac{(m_{0}^{2}/q^{2}%
)}{R_{q}}+O(1/R_{q}^{2})\right]  L_{[\alpha}N_{\beta]},
\end{equation}
and so, to first order, a free fall observer views the Coulomb field as a
$1/R_{q}$ wave zone radiation field.

\section{SUMMARY}

Classical radiation from an accelerating charge has been reviewed in Part I,
and falls within the framework of standard Maxwell theory. Part II treats the
reciprocal question of accelerated observers detecting radiation from a static
charge. Rohrlich and Mould have shown that such radiation is detectable. Their
answer has been affirmed here by studying a freely falling observer in the
static Reissner-Nordstr\"{o}m spacetime. The topic of radiation reaction has
been omitted, but there exists voluminous literature about the Lorentz-Dirac
equation. Within the citations at the end of this work, the interested reader
can find additional references to all related topics.

\appendix

\section{REISSNER-NORDSTR\"{O}M SPACETIME}

The RN metric in curvature coordinates has the standard form
\begin{equation}
ds_{\text{RN}}^{2}=(1-A_{q})dt^{2}-(1-A_{q})^{-1}dr^{2}-r^{2}d\Omega^{2}
\label{rnt_met}%
\end{equation}
where $A_{q}=2m_{0}/r-q^{2}/r^{2}$ and $d\Omega^{2}=d\vartheta^{2}+\sin
^{2}\vartheta\ d\varphi^{2}$.

Basis vectors, which span the RN metric and coincide with a Minkowski frame in
the $r\rightarrow\infty$ limit, are
\begin{subequations}
\begin{align}
\hat{t}_{\alpha}dx^{\alpha}  &  =(1-A_{q})^{1/2}dt,\label{t}\\
\hat{r}_{\alpha}dx^{\alpha}  &  =(1-A_{q})^{-1/2}dr,\label{r}\\
\hat{\theta}_{\alpha}dx^{\alpha}  &  =rd\vartheta,\label{theta}\\
\hat{\varphi}_{\alpha}dx^{\alpha}  &  =r\text{sin}\vartheta d\varphi,
\label{phi}%
\end{align}
with
\end{subequations}
\[
g_{\alpha\beta}^{\text{RN}}=\hat{t}_{\alpha}\hat{t}_{\beta}-\hat{r}_{\alpha
}\hat{r}_{\beta}-\hat{\theta}_{\alpha}\hat{\theta}_{\beta}-\hat{\varphi
}_{\alpha}\hat{\varphi}_{\beta}.
\]
This frame spans the RN metric outside the trapped surface located at
$1-A_{q}=0$.

The static Coulomb field has Maxwell tensor
\begin{equation}
F_{\alpha\beta}^{\text{RN}}=-(q/r^{2})(\hat{t}_{\alpha}\hat{r}_{\beta}-\hat
{r}_{\alpha}\hat{t}_{\beta}) \label{f_ten}%
\end{equation}
and trace-free Ricci tensor
\begin{equation}
R_{\alpha\beta}^{\text{RN}}=-(q^{2}/r^{4})(\hat{t}_{\alpha}\hat{t}_{\beta
}-\hat{r}_{\alpha}\hat{r}_{\beta}+\hat{\theta}_{\alpha}\hat{\theta}_{\beta
}+\hat{\varphi}_{\alpha}\hat{\varphi}_{\beta}). \label{Ricci-stat-tet}%
\end{equation}
Killing observers $k^{\beta}\partial_{\beta}=(1-A_{q})^{-1/2}\partial_{t} $
find a radial electric field $E_{\alpha}=F_{\alpha\beta}k^{\beta}%
=(q/r^{2})\delta_{\alpha}^{(r)}.$ The RN charge is given by the integral of
$F_{\alpha\beta}$ over any closed $t=const$, $r=const$, two-surface $S^{2}$
beyond the trapped surface:
\[%
{\displaystyle\oint\limits_{S^{2}}}
F_{\text{RN}}^{\alpha\beta}\ dS_{\alpha\beta}=4\pi q.
\]

\section{FREE FALL FRAME}

For observers falling from rest at infinity along radial geodesics, the RN
metric can be written in terms of the free fall proper time $\tau$
\cite{Dor99} where $dt=d\tau-A_{q}^{1/2}(1-A_{q})^{-1}dr$ with $A_{q}%
=2m_{0}/r-q^{2}/r^{2}$.%
\begin{equation}
ds_{\text{RN-ff}}^{2}=(1-A_{q})d\tau^{2}-2A_{q}^{1/2}d\tau dr-dr^{2}%
-r^{2}d\Omega^{2}. \label{rnff-met}%
\end{equation}
$g^{\text{RN-ff}}$ is spanned by%
\begin{align*}
\hat{\tau}_{\alpha}dx^{\alpha}  &  =(1-A_{q})^{1/2}d\tau-\left(  \frac{A_{q}%
}{1-A_{q}}\right)  ^{1/2}dr,\text{ \ }\hat{\tau}^{\alpha}\partial_{\alpha
}=(1-A_{q})^{-1/2}\partial_{\tau},\\
\hat{r}_{\alpha}dx^{\alpha}  &  =(1-A_{q})^{-1/2}dr,\text{ \ }\hat{r}^{\alpha
}\partial_{\alpha}=-\left[  \frac{1-A_{q}}{(1-A_{q})^{1/2}}\right]
\partial_{r}-\left(  \frac{A_{q}}{1-A_{q}}\right)  ^{1/2}\partial_{\tau},\\
\hat{\vartheta}_{\alpha}dx^{\alpha}  &  =rd\vartheta,\text{ \ \ \ \ \ \ \ }%
\hat{\varphi}_{\alpha}dx^{\alpha}=\sin\vartheta d\varphi
\end{align*}
such that
\begin{equation}
g_{\alpha\beta}^{\text{RN-ff}}=\hat{\tau}_{\alpha}\hat{\tau}_{\beta}-\hat
{r}_{\alpha}\hat{r}_{\beta}-\hat{\vartheta}_{\alpha}\hat{\vartheta}_{\beta
}-\hat{\varphi}_{\alpha}\hat{\varphi}_{\beta}.
\end{equation}
The radial free fall geodesics along radial lines\ $\Re$ have unit tangent
\begin{equation}
\hat{v}^{\alpha}\partial_{\alpha}=\partial_{\tau}-A_{q}^{1/2}\partial
_{r},\text{ \ \ }\hat{v}_{\alpha}dx^{\alpha}=d\tau\label{v-tan}%
\end{equation}
which satisfy $\hat{v}_{\ ;\beta}^{\alpha}\hat{v}^{\beta}=0$. This can be
written as
\begin{equation}
\frac{d\hat{v}^{\alpha}}{d\tau\ }+\Gamma_{\beta\nu}^{\alpha}\hat{v}^{\beta
}\hat{v}^{\nu}=0, \label{geod}%
\end{equation}
with radial acceleration
\begin{equation}
\ddot{r}=-\frac{1}{r^{2}}(m_{0}-q^{2}/r). \label{r-dot-dot}%
\end{equation}
Electromagnetic energy diminishes the gravitational effect of $m_{0}$.

A null frame [$L,N,M,\bar{M}$] along $\Re$ is constructed from the orthonormal
frame. $L$ is given by
\begin{align}
L_{\alpha}dx^{\alpha}  &  =\left[  \frac{(1-A_{q})^{1/2}}{1+A_{q}^{1/2}%
}\right]  d\tau-(1-A_{q})^{-1/2}dr,\label{El}\\
dR_{q}  &  =(1-A_{q})^{-1/2}dr,\nonumber
\end{align}
where $R_{q}$ is distance from the observer along the outgoing null direction.%
\begin{align}
R_{q}  &  =%
{\displaystyle\int\limits_{0}^{r}}
[1-2m_{0}/r^{\prime}+q^{2}/(r^{\prime})^{2}]^{-1/2}dr^{\prime}\label{R-q}\\
&  =(r^{2}-2m_{0}r+q^{2})^{1/2}+m_{0}\ln[r-m_{0}+(r^{2}-2m_{0}r+q^{2}%
)^{1/2}]\nonumber\\
&  -q-m_{0}\ln(q-m_{0})\nonumber
\end{align}
with constraint $m_{0}^{2}>q^{2}$. The other null vectors are%
\begin{align}
2N_{\alpha}dx^{\alpha}  &  =(1-A_{q})^{1/2}[(1+A_{q}^{1/2})d\tau-dr],
\label{En}\\
M_{\alpha}dx^{\alpha}  &  =-(r/\sqrt{2})(d\vartheta+i\sin\vartheta d\varphi).
\label{Em}%
\end{align}

\textbf{ACKNOWLEDGEMENT\ }\newline We thank Fritz Rohrlich for enlightening correspondence.

\end{document}